\documentclass[aps,prd,nofootinbib,twocolumn,superscriptaddress,amsmath,twoside,amssymbepsfig,showpacs]{revtex4-1}
\usepackage{graphicx}
\usepackage{dcolumn}
\usepackage{bm}
\usepackage{natbib}
\usepackage{color}
 \usepackage[normalem]{ulem}

\fontencoding{T1}

\newcommand{\aap}{{Astron.~Astrophys.}}

\newcommand{\apjs}{{Astrophys.~J.~Supp.}}

\newcommand{\mnras}{{Mon.~Not.~R.~Astron.~Soc.}}
\newcommand{\jcap}{{J.~Cosmol.~Astropart.~Phys.}}

\newcommand{\be}{\begin{equation}}
\newcommand{\ee}{\end{equation}}
\newcommand{\ba}{\begin{eqnarray}}
\newcommand{\ea}{\end{eqnarray}}

\newcommand{\WMAP}{{\slshape WMAP~}}
\newcommand{\LCDM}{$ \Lambda $CDM~}

\newcommand{\LCDMc}{$ \Lambda $CDM}
\newcommand{\Planck}{{\slshape Planck~}}

\begin{document}
\title{Bayesian evidence of non-standard inflation:\\
 Isocurvature perturbations and running spectral index}
\author {Tommaso Giannantonio}
\email{t.giannantonio at ast.cam.ac.uk}
\affiliation{Kavli Institute for Cosmology, Institute of Astronomy, University of Cambridge, Madingley Road, Cambridge CB3 0HA, UK}
\affiliation{Centre for Theoretical Cosmology, DAMTP, University of Cambridge, Wilberforce Road, Cambridge CB3 0WA, UK}
\affiliation{Ludwig-Maximilians-Universit\"at M\"unchen, Universit\"ats-Sternwarte M\"unchen, Scheinerstr. 1, D-81679 M\"unchen, Germany}
\author {Eiichiro Komatsu}
\affiliation {Max-Planck-Institut f\"{u}r Astrophysik,
Karl-Schwarzschild-Str. 1, D-85748 Garching, Germany}
\affiliation{Kavli Institute for the Physics and Mathematics of the
Universe, Todai Institutes for Advanced Study, the University of Tokyo,
Kashiwa, Japan 277-8583 (Kavli IPMU, WPI)}

\begin {abstract}
Bayesian model comparison penalizes models with more free
parameters that are allowed to vary over a wide range, and thus
offers the most robust method to decide whether some given data
require new parameters. In this paper, we ask a simple question: do
 current cosmological data require extensions of the simplest
single-field inflation models? Specifically, we calculate the Bayesian
evidence of a totally anti-correlated isocurvature perturbation and a
running spectral index of the scalar curvature perturbation. These
parameters are motivated by recent claims that the observed temperature
anisotropy of the cosmic microwave background on large angular
scales is too low to be compatible with the simplest inflation models.
Both a subdominant, anti-correlated cold dark matter isocurvature
component and a negative running index succeed in lowering the
large-scale temperature power spectrum. We show that
the introduction of  isocurvature perturbations is disfavored,
whereas that of the running spectral index is only moderately favored,
even when the BICEP2 data are included in the analysis without any
foreground subtraction.
\end {abstract}
\maketitle
\section {Introduction} \label {sec:intro}
Suppose that we wish to decide whether some data require
the addition of a new parameter to a model. We may compare the
logarithms of the likelihood values evaluated at the best-fit 
parameters. For example, the conventional $\chi^2$ method uses
$\Delta\chi^2\equiv -2\ln({\cal L}_1/{\cal L}_2)$. The obvious 
problem of this approach is that the addition of a new parameter is
guaranteed to improve the fit, yielding a smaller $\chi^2$ value. But
then, what does $\Delta\chi^2$ mean when we find, say,
$\Delta\chi^2=-7$ by adding one more parameter? Do the data require such
a parameter? 

To address this issue, some criteria for comparing models have been
discussed in the literature. The Akaike information criterion (AIC;
\cite{akaike:1974}) 
and the Bayesian information criterion (BIC; \cite{schwarz:1978}) penalize models with
more parameters by adding to $\chi^2$ a term proportional to the number
of parameters. These criteria penalize all parameters
equally regardless of predictability. For example,
consider two parameters, one being allowed to vary from $-1$
to 1, and the other from 0 to $10^{10}$. While AIC and BIC penalize
both parameters equally, a more sensible criterion should penalize the
latter more strongly. 

In this paper, we shall apply \emph{Bayesian model comparison}
\cite{jeffreys:1961} to test whether extensions of the simplest
inflation models are required by the current cosmological data. The
Bayesian model comparison penalizes models with more free parameters
that are allowed to vary over a wide range. Specifically, we compute the
Bayesian evidence, ${\cal Z}$, defined by
\begin{equation}
 {\cal Z}\equiv \int d^N\theta~{\cal L}({\rm
  data}|{\bm\theta})P({\bm\theta }),
\label{eq:evidence}
\end{equation}
where ${\cal L}({\rm data}|{\bm\theta})$ is the likelihood of the data given
the model parameters ${\bm\theta}$, and $P({\bm\theta })$ is the prior
probability. We then compare two models by computing the logarithm of
the ratio of their evidences, $\ln B\equiv \ln({\cal
Z}_1/{\cal Z}_2)$. Since the prior probability is normalized as $\int
d^N\theta~P({\bm\theta})=1$,  $P({\bm\theta})$ at a
given set of $\bm\theta$ becomes small when a model
contains more parameters varying over a wide range. This gives that
model a small ${\cal Z}$, hence penalizing it more strongly.
 The factor $\ln B$ can be interpreted as the mathematical odds between
the models given the data, which can also be expressed heuristically
using the so-called ``Jeffrey's scale'', according to which the evidence
for (or against) a model is said to be weak, moderate, and strong if
$\ln B > 1$, $2.5$, and  $5$, respectively
\cite{2008ConPh..49...71T}. We shall adopt Jeffrey's scale throughout
this paper.

Why consider extensions of the simplest inflation
models? Here, the ``simplest inflation models'' refer to inflation
models driven by a single scalar field with a simple potential yielding
approximately a power-law power spectrum of the scalar curvature perturbation.

A detection of isocurvature modes of any form would rule
out all single-field inflation models. Moreover, a
detection of a cold dark matter (CDM) isocurvature mode
would shed light on the nature of CDM, e.g., axions 
\cite{kolb/turner:TEU}.

Given that the measured deviation of the scalar curvature power
spectrum from scale invariance is $1-n_s \simeq 0.04$
\cite{hinshaw/etal:2013,2013arXiv1303.5076P}, the running spectral
index, $\rho_s\equiv dn_s/d\ln k$, is typically of order $(1-n_s)^2={\cal
O}(10^{-3})$; however, larger values are possible if the third
derivative of the potential of a scalar field driving inflation is large
\cite{lyth/riotto:1999}. Thus, a large running index of order $10^{-2}$
necessarily requires a new energy scale in the potential (either
in the kinetic term of the field \cite{chung/shiu/trodden:2003} or in the
initial vacuum state \cite{2014arXiv1403.6099A}), making the
models more complicated.

A motivation to consider these extensions of the simplest single-field
inflation models comes from the observational data of the cosmic
microwave background (CMB). The \Planck collaboration claims that the
CMB temperature power spectrum data that they obtain at low multipoles
are too low to be compatible with the best-fit power-law ($\rho_s=0$)
adiabatic curvature perturbation spectrum
\cite{2013arXiv1303.5076P}. Both a negative running index and a nearly
scale-invariant CDM isocurvature component that is anti-correlated
with the curvature perturbation can lower the low-multipole power,
reducing this apparent ``tension'' in the \Planck temperature data
\cite{2013arXiv1303.5082P}.

This tension is exacerbated \cite{2014arXiv1404.0373S}, if a significant fraction of the B-mode
polarization detected at degree angular scales by the BICEP2 collaboration
\cite{2014arXiv1403.3985B} originates from the primordial, nearly
scale-invariant gravitational waves generated during inflation, as such
gravitational waves add extra power to the  temperature power
spectrum at low multipoles \cite{starobinsky:1985}. Then, do the \Planck
and BICEP2 data require either a negative running index or an anti-correlated
CDM isocurvature perturbation? This is the question that we shall address in
this paper using Bayesian model comparison.

Ref.~\cite{2014arXiv1403.5922A} computed the Bayesian evidence
of a running index, showing that evidence for running is
insignificant. Our results differ from theirs because of the choice
of the data set and the prior probability on the amplitude of
gravitational waves. 

Refs.~\cite{2014JCAP...06..061H,
2014arXiv1404.0360H,2014arXiv1405.2012H, 2014arXiv1406.3243M} computed
$\Delta\chi^2$ for inflation models which produce modifications of the
primordial power spectrum at small wavenumbers, but did not perform a
Bayesian model comparison. Ref.~\cite{2014arXiv1404.2175K} computed
$\Delta\chi^2$ for isocurvature perturbations, but did not perform a
Bayesian model comparison. Thus, they were unable to conclude whether the
data require such extensions of the simple inflation models.

The structure of this paper is as follows: 
We describe the models in Section
\ref{sec:theory}, and  present the data sets we use 
and the analysis method in Section \ref{sec:data}. We describe our results
in Section \ref{sec:results}, and conclude in Section \ref{sec:conclusion}.

\section {Models} \label{sec:theory}

\begin{figure}
\includegraphics[width=\linewidth]{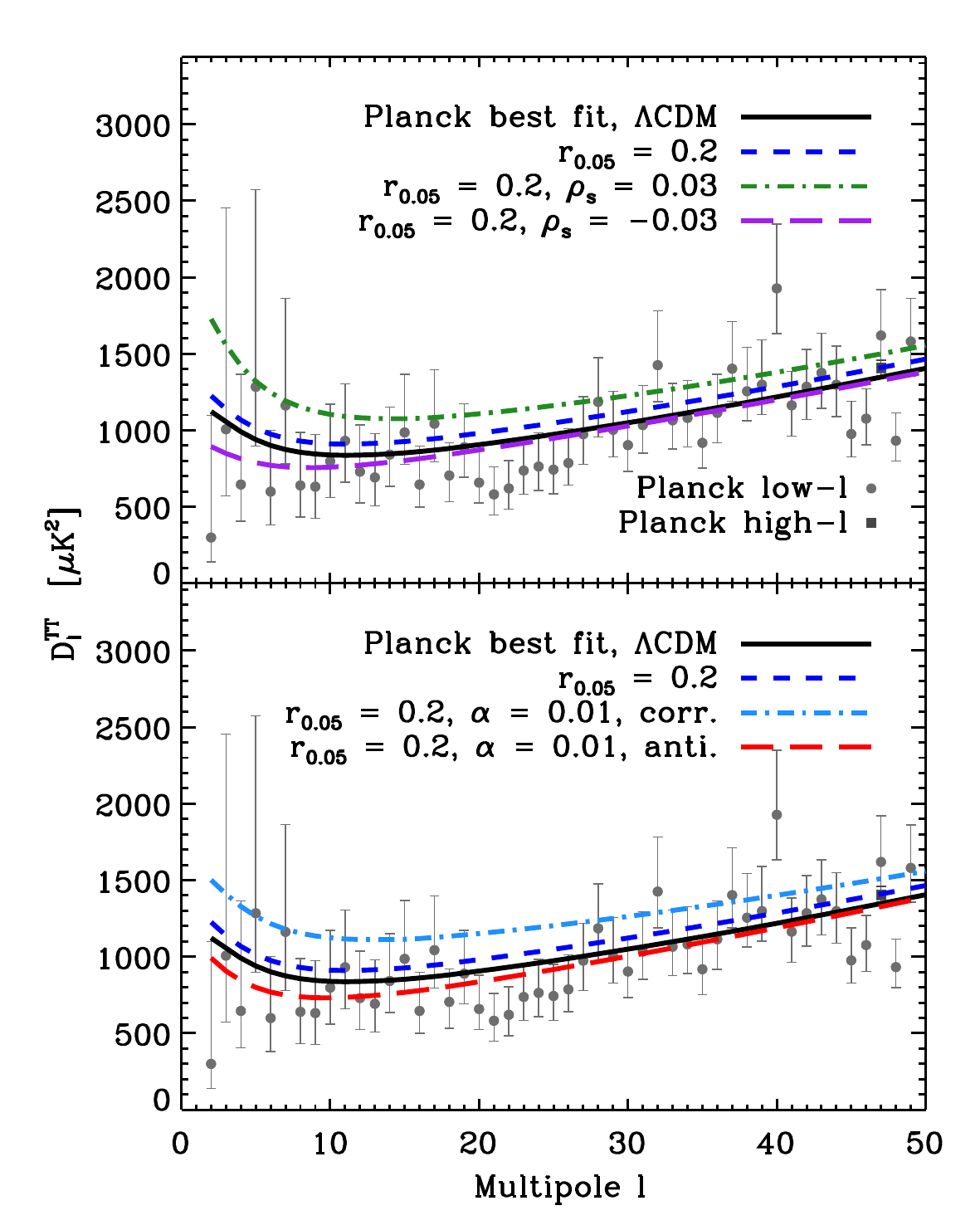}
\caption{Comparison of the \LCDM and extended
 models. In both panels, the solid lines show the
scalar CMB power spectrum of the
 six-parameter $\Lambda$CDM model, while the short-dashed lines show the
 sum of the solid lines and the tensor power spectrum with
 a tensor-to-scalar ratio of
 $r_{0.05}=0.2$. The symbols with error bars show the \Planck
 measurements \cite{2013arXiv1303.5075P}.
 (Top panel:) 
The long-dashed and dot-dashed lines show the sum of the tensor power
 spectrum and the scalar power spectrum with negative and positive
 running indices,  respectively, with $|\rho_s|=0.03$.
(Bottom panel:)
The long-dashed and dot-dashed lines show
 the sum of the short-dashed line and totally anti-correlated and
 correlated CDM isocurvature components, respectively, with an
 isocurvature-to-curvature ratio of $\alpha=0.01$. 
}
\label{fig:theory}
\end{figure}

\subsection{Model I: Running scalar spectral index}

We write the scalar curvature power spectrum as
\begin{align}
\mathcal{P}_{\cal R} (k) = A_s \, \bar k^{n_s - 1 + \frac{1}{2} \rho_s \ln \bar k} \, ,
\label{eq:scalar}
\end{align}
where $n_s$ and $\rho_s$ are the scalar spectral index and
its running, respectively, and $\bar k\equiv k/(0.05~{\rm Mpc}^{-1})$ is
the normalized wavenumber. The tensor power spectrum is
\begin{align}
\mathcal{P}_{h} (k) = r_{0.05} \, A_s\bar k^{-r_{0.05}/8} \, , 
\label{eq:tensor}
\end{align}
where $r_{0.05}$ is the tensor-to-scalar ratio defined at
$k=0.05~{\rm Mpc}^{-1}$.

In the top panel of Fig.~\ref{fig:theory} we compare the temperature power
spectrum data, $D_l\equiv l(l+1)C_l/(2\pi)$, measured by \Planck
\cite{2013arXiv1303.5075P} with three representative models. The solid 
line shows the best-fit six-parameter adiabatic $\Lambda$CDM model with
$\rho_s=0$ and $r_{0.05}=0$. The short-dashed  line is the sum of the
solid  line and the tensor temperature power spectrum with
$r_{0.05}=0.2$, showing how adding the tensor power spectrum with
the tensor-to-scalar ratio suggested by the BICEP2 data (without
foreground subtraction) 
exacerbates the tension between the model and the \Planck temperature
data. The long-dashed line has $r_{0.05}=0.2$ and a negative running
index of $\rho_s=-0.03$, which brings the model back in agreement with
the data. The dot-dashed line has a positive running index, yielding a
bad fit.

\subsection{Model II: CDM isocurvature}

When we study an isocurvature component, we use Eq.~(\ref{eq:scalar}) for
the scalar curvature power spectrum with $\rho_s\equiv 0$.
We continue to use the same tensor power spectrum as Eq.~(\ref{eq:tensor}).
We write the power spectrum of an isocurvature component, ${\cal S}$, as
\begin{align}
\mathcal{P}_{\mathcal{S}} (k) 
= \alpha A_s \, \bar k^{n_{\mathrm{iso}} - 1}, 
\end{align}
where $n_{\mathrm{iso}}$ is the corresponding spectral
index, and $\alpha$ is the isocurvature-to-curvature power ratio at
$k=0.05~{\rm Mpc}^{-1}$. 
We shall assume that ${\cal R}$ and ${\cal S}$ are totally
anti-correlated (or correlated) throughout this paper.
We thus write the cross-correlation power spectrum between ${\cal R}$
and ${\cal S}$ as
\begin{align}
\mathcal{P}_{\mathcal{RS}} (k) 
=\pm \sqrt{P_{\cal R}(k)P_{\cal S}(k)}\, .
\end{align}
To minimize the number of parameters, we set
$n_{\rm iso}=n_s$.

In the lower panel of Fig.~\ref{fig:theory}, the solid
line shows the best-fit six-parameter adiabatic $\Lambda$CDM model with
$\alpha=0$ and $r_{0.05}=0$. The short-dashed line is the sum of the
black line and the tensor temperature power spectrum with
$r_{0.05}=0.2$, again showing that the BICEP2 data without
foreground subtraction exacerbate the tension. The long-dashed line has
$r_{0.05}=0.2$ and 
a totally anti-correlated 
isocurvature component 
with $\alpha=0.01$, 
which brings the model back in agreement with 
the data. The dot-dashed line has a totally correlated isocurvature
component with $\alpha=0.01$, yielding a bad fit.

\section {Data and analysis method} \label{sec:data}

We use the \Planck temperature power spectrum from the 2013 public
release \cite{2013arXiv1303.5075P}, with the addition of the \WMAP
9-year polarization data \cite{2013ApJS..208...19H} as combined in the
default analysis by the \Planck collaboration, as well as
the B-mode polarization power spectrum released by the BICEP2 collaboration \cite{2014arXiv1403.3985B}.

We also include 
a suite of baryon acoustic oscillation (BAO) distance scale measurements
by the BOSS and 6dF collaborations, using the BOSS data release 9 (DR9)
measurement at $z \simeq 0.57$ \cite{2012MNRAS.427.3435A}, the DR7
measurement at $z \simeq 0.35$ \cite{2012MNRAS.427.2132P}, and 6dF
result at $z \simeq 0.1 $ \cite{2011MNRAS.416.3017B}. We do not use any
supernovae or $H_0$ data.

\begin{table} 
\begin{ruledtabular}
\begin{tabular}{ccc} 
Parameter           &      Description      &     Priors   \\
\hline
$\omega_b \equiv \Omega_b h^2$   & baryonic energy density & [0.020, 0.025] \\
$\omega_c \equiv \Omega_c h^2$   & dark matter energy density & [0.080, 0.16] \\
$100 \, \vartheta $ & sound horizon at last scattering & [1.034, 1.045] \\
$\tau$ & optical depth & [0.05, 0.18] \\
$n_s$ & scalar spectral index & [0.90, 1.05] \\
$\log (10^{10} A_s)$ & scalar amplitude & [2.9, 3.35] \\
\hline
$r_{0.05}$      & tensor-to-scalar ratio   &  [0.0, 1.0]\\
$\alpha$   & isocurvature-to-curvature ratio & [0.0, 1.0] \\
$\rho_s$   & scalar running spectral index & [$-0.1$, 0.1] \\
\end{tabular}
\end{ruledtabular}
\caption{Parameters considered and prior ranges. In addition to these, all standard \Planck nuisance parameters are left free and marginalized over.}
\label{tab:params}
\end{table}

We perform a Bayesian Monte Carlo exploration of the
parameter space, 
using nested sampling as implemented in the public code
\textsc{Multinest} \cite{2009MNRAS.398.1601F,2013arXiv1306.2144F}, used
as an alternative sampler within the \textsc{Cosmomc/Camb} code
\cite{Lewis:2000a, Lewis:2002a}. This method allows us to directly
estimate the Bayesian evidence of each model and its uncertainties, and
to compare them.  

We let the parameters vary freely within the ranges described in
Table~\ref{tab:params}. As the nested sampling algorithm starts from
uniform sampling over the whole parameter space,  it is desirable to choose
tight prior ranges such that the sampling is efficient.
 We thus choose a prior distribution  for the standard \LCDM
 parameters that is narrow, while being sufficiently broad so that the posterior
 likelihood of the six parameters is zero near the edges of
 the prior.

 The prior
 distribution of the new parameters, i.e., $r_{0.05}$, $\alpha$, and
 $\rho_s$, is chosen such that the power of tensor or isocurvature
 perturbations does not exceed that of the scalar curvature
 perturbation ($r_{0.05}\in [0,1]$ and $\alpha\in [0,1]$), and that the
 running spectral index is not too much bigger than $|1-n_s|$ ($\rho_s\in
 [-0.1,0.1]$). These prior distributions make physical sense
 and are compatible with expectations from
 inflation.

In addition to the parameters shown in Table~\ref{tab:params}, we
include the entire list of the standard \Planck nuisance parameters,
over which we marginalize. As in the standard \Planck analysis, we
account for massive neutrinos with a total mass fixed at $\sum m_{\nu}
= 60 $~meV.

\section {Results}  \label{sec:results}
\squeezetable
\begin{table*}
\begin{ruledtabular}
\begin{tabular}{cccccc}
Data         & Model                  &        Best fits                &     Best-fit $\chi^2$   &   $\Delta \chi^2$ w.r.t. \LCDM   &  $\Delta \chi^2$ w.r.t. $r$\LCDMc   \\
\hline
\Planck + WP & \LCDM                  &    ---                           &  $ 9804.1 $       & ---         &     $ 0.0 $    \\
+ BAO        & + $r_{0.05}$            & $ r_{0.05} = 5.6 \cdot 10^{-4}$   &  $ 9804.1 $       & $ 0.0 $   &     ---    \\
             & + $\alpha$             & $ \alpha = 7.1 \cdot 10^{-4}$    &  $ 9803.2 $       & $ -0.9 $    &   $ -0.9 $   \\
             & + $\rho_s$             & $ \rho_s =  -0.012 $             &  $ 9802.8 $       & $ -1.4 $    &   $ -1.4 $   \\
             & + $r_{0.05}$ + $\alpha$ & $ r_{0.05} = 1.7 \cdot 10^{-4} $;  &  $ 9803.2 $       & $ -0.9 $   &    $ -0.9 $     \\
             &                        & $ \alpha = 6.5 \cdot 10^{-4} $    &                           &                   &                    \\
             & + $r_{0.05}$ + $\rho_s$ & $ r_{0.05} = 0.0020   $;          & $ 9802.8 $       & $ -1.4 $    &   $ -1.4 $   \\
             &                        & $ \rho_s = -0.013 $ &                          &                    &                    \\
\hline
\Planck + WP + & \LCDM                &    ---                           & $ 9860.2 $        & ---  &  $40.2$  \\
BICEP2 B-mode& + $r_{0.05}$            & $ r_{0.05} = 0.16 $               &  $ 9820.0 $       & $ -40.2 $  &   ---  \\
+ BAO        & + $\alpha$             & $ \alpha = 1.1 \cdot 10^{-3} $   &  $ 9858.8 $       & $ -1.3 $ &  $38.8$ \\
             & + $\rho_s$             & $ \rho_s =  -0.015 $        &  $ 9858.1 $             & $ -2.0 $ &  $ 38.1 $ \\
             & + $r_{0.05}$ + $\alpha$ & $ r_{0.05} = 0.17 $;              &  $ 9815.7 $       & $ -44.4 $  & $-4.2$ \\
             &                        & $ \alpha = 0.0036 $              &                           &                   &             \\
             & + $r_{0.05}$ + $\rho_s$ & $ r_{0.05} = 0.19 $;              &  $ 9812.7 $         & $ -47.4 $   & $-7.2$\\
             &                        & $ \rho_s = -0.032 $              &                            &                   &               \\
\end{tabular}
\end{ruledtabular}
\caption{Frequentist analysis results.}
\label{tab:resultsfreq}
\end{table*}

\squeezetable
\begin{table*}
\begin{ruledtabular}
\begin{tabular}{cccccccc}
Data         & Model                  &         95\% c.l. posteriors    &   $\ln (Z)$            &   $\ln B = \Delta \ln Z$ &    Jeffrey's scale & $\ln B$ w.r.t. $r$\LCDMc &  Jeffrey's scale \\
\hline
\Planck + WP & \LCDM                  &    ---                          & $ -4940.94  \pm 0.05 $ &  ---                     & ---                &   $2.82 \pm 0.06 $      &   moderate in favor \\
+ BAO        & + $r_{0.05}$            & $ r_{0.05} \in [0, 0.12] $       & $ -4943.76  \pm 0.03 $ &  $-2.82 \pm 0.06 $       & moderate against    &   ---       &  ---          \\
             & + $\alpha$             & $ \alpha \in [0, 0.0073] $      & $ -4945.71  \pm 0.04 $ &  $-4.77 \pm 0.06 $       & moderate against    &   $-1.95 \pm 0.05 $    &    weak against      \\
             & + $\rho_s$             & $ \rho_s \in [-0.031, 0.0033] $ & $ -4941.89  \pm 0.03 $ &   $-0.95 \pm 0.06 $       & inconclusive     &    $ 1.87 \pm 0.04 $      &  weak in favor   \\
             & + $r_{0.05}$ + $\alpha$ & $ r_{0.05} \in [0, 0.19] $;      & $ -4947.66  \pm 0.07 $ &  $-6.72 \pm 0.09 $       & strong   against   &    $-3.90 \pm 0.08 $     &    moderate against      \\
             &                        & $ \alpha \in [0, 0.010] $      &                        &                          &                    &                   &                    \\
             & + $r_{0.05}$ + $\rho_s$ & $ r_{0.05} \in [0, 0.24] $;      & $ -4943.65  \pm 0.04 $ &  $-2.71 \pm 0.06 $       & moderate   against  &  $0.11 \pm 0.05 $       &  inconclusive        \\
             &                        & $ \rho_s \in [-0.043, -0.00035] $ &                       &                          &                    &                    &                    \\
\hline
\Planck + WP + & \LCDM                  &    ---                        & $ -4969.07  \pm 0.01 $ &  ---                     & ---                &   $-17.39 \pm 0.04 $      &  strong against  \\
BICEP2 B-mode& + $r_{0.05}$            & $ r_{0.05} \in [0.093, 0.23] $   & $ -4951.68  \pm 0.04 $ &  $\mathbf{17.39 \pm 0.04} $       & \textbf{strong in favor}   &    ---     &   ---   \\
+ BAO        & + $\alpha$             & $ \alpha \in [0, 0.0079] $      & $ -4973.42  \pm 0.10 $ &  $-4.35 \pm 0.10 $       & moderate against     &   $-21.74 \pm 0.11 $       &  strong against   \\
             & + $\rho_s$             & $ \rho_s \in [-0.035, 0.00044] $      & $ -4969.58  \pm 0.02 $ &  $-0.51 \pm 0.02 $       & inconclusive  &   $-17.90 \pm 0.04 $       &  strong against   \\
             & + $r_{0.05}$ + $\alpha$ & $ r_{0.05} \in [0.11, 0.26] $;   & $ -4953.94  \pm 0.01 $ &  $15.13 \pm 0.01 $       & strong in favor  &    $\mathbf{-2.26 \pm 0.04} $      & \textbf{weak against}     \\
             &                        & $ \alpha \in [0, 0.013] $       &                        &                          &                    &                   &             \\
             & + $r_{0.05}$ + $\rho_s$ & $ r_{0.05} \in [0.12, 0.27] $;     & $ -4949.16  \pm 0.03 $ &  $19.91 \pm 0.03 $       & strong in favor  &   $\mathbf{2.52 \pm 0.05} $      &  \textbf{moderate}   \\
             &                        & $ \rho_s \in [-0.050, 0.011] $     &                        &                          &                    &                   &       \textbf{in favor}         \\
\end{tabular}
\end{ruledtabular}
\caption{Bayesian analysis results.}
\label{tab:results}
\end{table*}

\subsection{Frequentist analysis: $\Delta\chi^2$}

Let us first show the results from the frequentist analysis
using the usual $\Delta\chi^2$ statistics. The sixth column of
Table~\ref{tab:resultsfreq} shows $\Delta\chi^2$ values between
\LCDMc+$r_{0.05}$ and the other models. Negative values indicate
a better fit over the former model. The first column shows the data
combinations. When the BICEP2 data are included, we find
$\Delta\chi^2=-7.2$ and $-4.2$ for the running spectral index and the
anti-correlated CDM isocurvature models, respectively.\footnote{Notice that, while we reproduce the best-fit values of Ref.~\cite{2014arXiv1404.2175K} for the anticorrelated isocurvature case, we find a smaller $\chi^2$ improvement than these authors: we find $\Delta\chi^2=-4.7$ when using their same settings, while they quote $-5.8$. After private communications, we have found that this discrepancy is due to numerical inaccuracies in the best-fit search of Ref.~\cite{2014arXiv1404.2175K}.}
The isocurvature mode
gives a smaller improvement because, while it reduces the low-multipole temperature power spectrum, it also reduces the power at $l \sim 300$ slightly, which is disfavored by the data.

Both models contain one more free
parameter than \LCDMc+$r_{0.05}$. While the $\Delta\chi^2$ values tell
us that introducing one more parameter improves the fit, they do not tell
us whether the data {\it require} such a parameter.

\subsection{Bayesian evidences}

Next, we show the results from the Bayesian analysis
using the logarithms of the evidence ratio, $\ln B$. The seventh column of
Table~\ref{tab:results} shows $\ln B$ values between
\LCDMc+$r_{0.05}$ and the other models. {\it Positive} values indicate
that the other models are favored over \LCDMc+$r_{0.05}$. When the BICEP2
data are included, we find 
$\ln B = 2.52$ and $-2.26$
 for the running spectral index and the CDM
isocurvature models, respectively. These results clearly show the power of 
Bayesian model comparison: despite an improved $\chi^2$, the
anti-correlated CDM isocurvature model is {\it disfavored} by the
data. The running spectral index model is still favored, and it is
``moderately favored'' according to Jeffrey's scale. We have also tested  the effect of changing the priors by reducing the assumed range on running by a factor of two to $\rho_s \in [-0.05, 0.05]$. We have found that in this case the result simply reflects the change in the prior volume: the Bayes factor grows by a factor of $\Delta \ln B \simeq \ln 2$ from $\ln B = 2.5$ to $\ln B = 3.1$. Furthermore, we have tried for the isocurvature case a uniform logarithmic prior: $\mathrm{Log}_{10} \alpha \in [-6, 0]$. We find that also in this case the model with isocurvature is not strongly favoured compared with the $r\Lambda$CDM case, as $ \ln B = 1.23 \pm 0.05 $, which is weakly favoured on Jeffrey's scale.
Broader choices of the logarithmic prior would further penalize the model, while narrower choices would be fine-tuned and would quickly exclude parts of the parameter space near the best-fit point.

\begin{figure}
\includegraphics[width=0.47\linewidth]{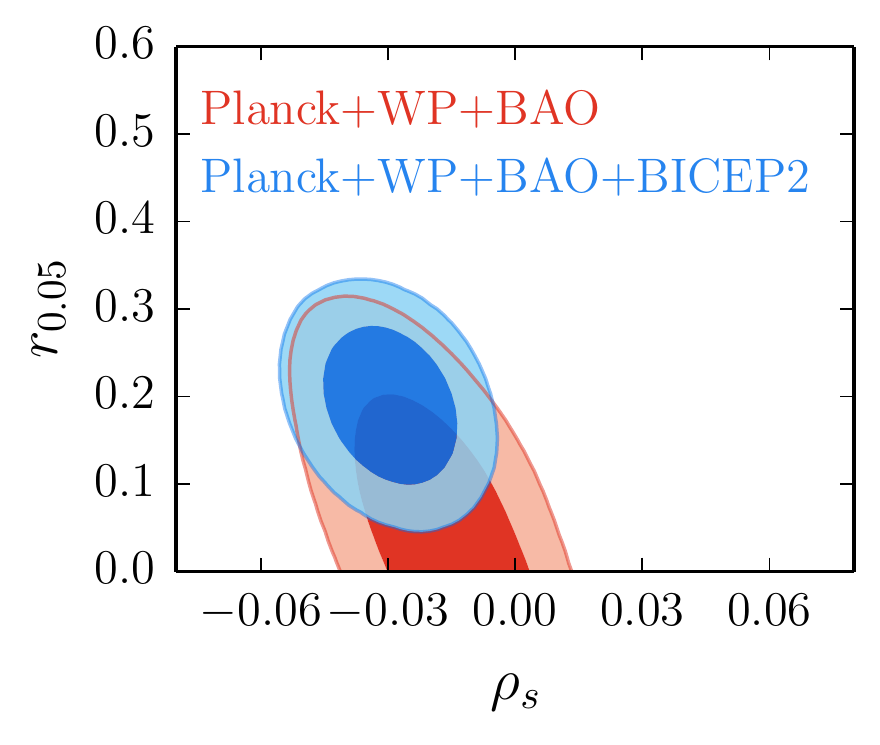}
\includegraphics[width=0.49\linewidth]{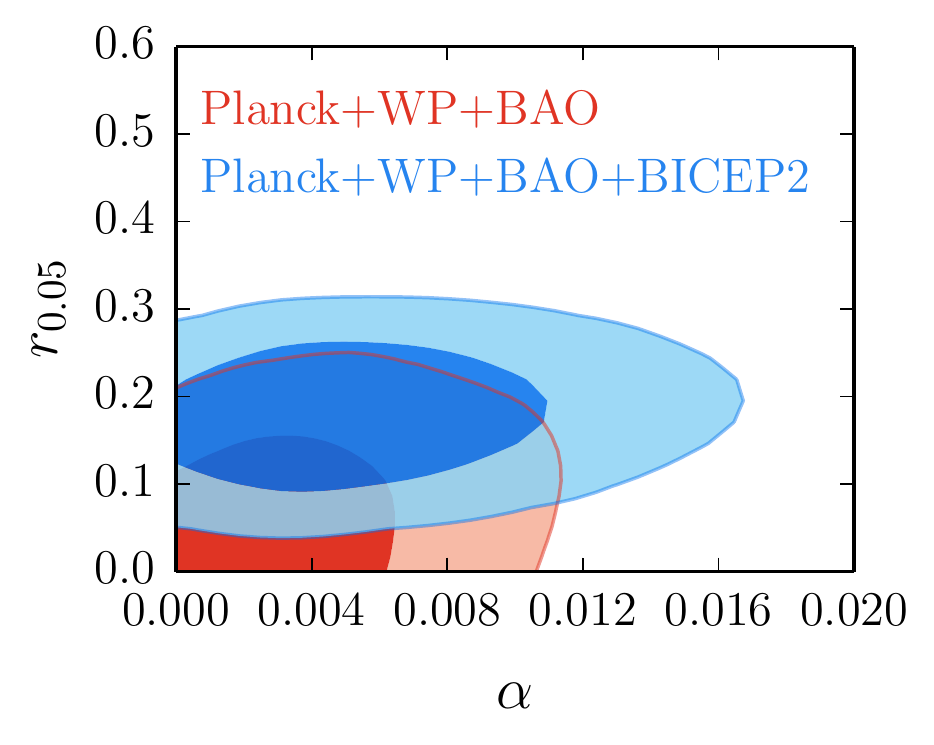}
\caption{Marginalized 2D posteriors on the tensor-scalar ratio, running, and isocurvature parameters.}
\label{fig:2Dplots}
\end{figure}

We show the marginalized 2D posteriors on the parameters of interest in Fig.~\ref{fig:2Dplots}, where we can see a visual confirmation of
the 95\% confidence intervals shown in the third column of
Table~\ref{tab:results}: the scalar running is favored at the $2\sigma$
level, while the amount of anti-correlated CDM isocurvature is
consistent with zero.

We have tested the stability of our results when
including the \Planck CMB lensing likelihood, removing BAOs, and using
${\cal P}_h\propto \bar k^0$ instead of $\bar k^{-r/8}$.
We find that the results are relatively robust, although the evidence in favor of running is reduced in some of these cases:
the addition of CMB lensing in particular reduces the evidence to $\ln B=1.8$,
which is ``weak'' on Jeffrey's scale. 
The \Planck collaboration also finds a reduced significance of a running index
when using the CMB lensing data \cite{2013arXiv1303.5076P}.

Our results change more significantly if the same method
of Ref.~\cite{2014arXiv1403.5922A} is used, where the posterior
likelihood of the tensor-to-scalar ratio obtained by the BICEP2 collaboration was used as a prior
instead of calculating the full BICEP2 likelihood for each model. If we
use their method, we reproduce their results, which show an even smaller
evidence ratio for the running spectral index model, $\ln B=1.1$. While
applying the BICEP2 posterior distribution on $r_{0.05}$ as a prior 
is reasonable when constraining the tensor amplitude only, the results will
be only approximately recovered if both $r_{0.05}$ and
$\rho_s$ are varied simultaneously. This is because the BICEP2 posterior was obtained for a model without running, so that any degeneracy between $r_{0.05}$ and $\rho_s$ will be missed if using this approach. We thus conclude that
Ref.~\cite{2014arXiv1403.5922A}  underestimated the evidence ratio for
the running spectral index model.

\section {Conclusions} \label{sec:conclusion}

There are at least three easy ways to reduce the apparent
``tension'' between the simplest inflation models with a tensor mode and
the current CMB data including \Planck and BICEP2. First, a sub-dominant
CDM isocurvature perturbation anti-correlated with the
dominant curvature perturbation \cite{2014arXiv1404.2175K, 2014arXiv1404.4976B}; second, a
negative running spectral index \cite{2014arXiv1403.3985B}; and third, a
modification of the large-scale primordial power spectrum
\cite{2014arXiv1403.5922A, 2014arXiv1404.2278B, 2014arXiv1404.2274F, 2014JCAP...06..061H, 2014arXiv1404.0360H,2014arXiv1405.2012H, 2014arXiv1406.3243M}. 

We have performed a Bayesian model comparison of the former two
extensions against the simplest inflation models. The
anti-correlated CDM isocurvature component reduces the CMB
temperature power spectrum at low multipoles, improving the
agreement with the tensor model with $r_{0.05}=0.2$ suggested by the BICEP2 data
without any foreground subtraction. Nonetheless, we have found that such
an improvement is Bayesianly {\it disfavored}, i.e., the data do not support
such an extension of the inflation model, despite that it gives an
improved $\chi^2$ by $\Delta\chi^2=-4.2$. This shows the power of the
Bayesian model comparison method.
While this result necessarily depends on the chosen prior on the amount
of isocurvature, i.e., $\alpha\in [0,1]$, this prior is physically
motivated, and there is little room for ambiguity on the prior choice. 

We have then tested a model with a running spectral index, as a negative
running can also reduce the temperature power spectrum at low
multipoles. We have found that a negative running spectral index is
moderately favored with the log evidence ratio of $\ln B=2.52$. 

Our results are derived assuming that there is no
foreground contamination in the BICEP2 data. Any foreground
contributions will lower $\ln B$,
 and thus the anti-correlated CDM isocurvature will be even more disfavored, and
the evidence for a negative running spectral index will likely turn to
be ``weak'' ($\ln B<2.5$). The BICEP2 collaboration
finds that the polarized dust emission could account for 30\% of the
measured B-mode power spectrum, while others argue that 100\% could
be accounted for by dust \cite{2014arXiv1405.5857M,
2014arXiv1405.7351F}. Therefore, we conclude that the current data
do not require these particular extensions of the simplest inflation models.

~

\section*{Acknowledgments}

We thank Grigor Aslanyan, Christian T. Byrnes, Richard Easther, Jussi
V\"{a}liviita and Jochen Weller for useful discussion,
Guillermo Ballesteros for comments on the prior,
and Toyokazu Sekiguchi for exchanging the results of his best-fit estimates.
range of the running spectral index. Numerical calculations were run on the Hydra supercomputer of the Max Planck Society in Garching, Germany. 

\appendix

\bibliographystyle{apsrev4-1}
\bibliography{ms}

\begin{thebibliography}{34}%
\makeatletter
\providecommand \@ifxundefined [1]{%
 \@ifx{#1\undefined}
}%
\providecommand \@ifnum [1]{%
 \ifnum #1\expandafter \@firstoftwo
 \else \expandafter \@secondoftwo
 \fi
}%
\providecommand \@ifx [1]{%
 \ifx #1\expandafter \@firstoftwo
 \else \expandafter \@secondoftwo
 \fi
}%
\providecommand \natexlab [1]{#1}%
\providecommand \enquote  [1]{``#1''}%
\providecommand \bibnamefont  [1]{#1}%
\providecommand \bibfnamefont [1]{#1}%
\providecommand \citenamefont [1]{#1}%
\providecommand \href@noop [0]{\@secondoftwo}%
\providecommand \href [0]{\begingroup \@sanitize@url \@href}%
\providecommand \@href[1]{\@@startlink{#1}\@@href}%
\providecommand \@@href[1]{\endgroup#1\@@endlink}%
\providecommand \@sanitize@url [0]{\catcode `\\12\catcode `\$12\catcode
  `\&12\catcode `\#12\catcode `\^12\catcode `\_12\catcode `\%12\relax}%
\providecommand \@@startlink[1]{}%
\providecommand \@@endlink[0]{}%
\providecommand \url  [0]{\begingroup\@sanitize@url \@url }%
\providecommand \@url [1]{\endgroup\@href {#1}{\urlprefix }}%
\providecommand \urlprefix  [0]{URL }%
\providecommand \Eprint [0]{\href }%
\providecommand \doibase [0]{http://dx.doi.org/}%
\providecommand \selectlanguage [0]{\@gobble}%
\providecommand \bibinfo  [0]{\@secondoftwo}%
\providecommand \bibfield  [0]{\@secondoftwo}%
\providecommand \translation [1]{[#1]}%
\providecommand \BibitemOpen [0]{}%
\providecommand \bibitemStop [0]{}%
\providecommand \bibitemNoStop [0]{.\EOS\space}%
\providecommand \EOS [0]{\spacefactor3000\relax}%
\providecommand \BibitemShut  [1]{\csname bibitem#1\endcsname}%
\let\auto@bib@innerbib\@empty
\bibitem [{\citenamefont {{Akaike}}(1974)}]{akaike:1974}%
  \BibitemOpen
  \bibfield  {author} {\bibinfo {author} {\bibfnamefont {H.}~\bibnamefont
  {{Akaike}}},\ }\href@noop {} {\bibfield  {journal} {\bibinfo  {journal} {IEEE
  Transactions on Automatic Control}\ }\textbf {\bibinfo {volume} {19}},\
  \bibinfo {pages} {716} (\bibinfo {year} {1974})}\BibitemShut {NoStop}%
\bibitem [{\citenamefont {{Schwarz}}(1978)}]{schwarz:1978}%
  \BibitemOpen
  \bibfield  {author} {\bibinfo {author} {\bibfnamefont {G.}~\bibnamefont
  {{Schwarz}}},\ }\href@noop {} {\bibfield  {journal} {\bibinfo  {journal}
  {Annals of Statistics}\ }\textbf {\bibinfo {volume} {6}},\ \bibinfo {pages}
  {461} (\bibinfo {year} {1978})}\BibitemShut {NoStop}%
\bibitem [{\citenamefont {{Jeffreys}}(1961)}]{jeffreys:1961}%
  \BibitemOpen
  \bibfield  {author} {\bibinfo {author} {\bibfnamefont {H.}~\bibnamefont
  {{Jeffreys}}},\ }\href@noop {} {\emph {\bibinfo {title} {Theory of
  Probability}}},\ \bibinfo {edition} {3rd}\ ed.\ (\bibinfo  {publisher}
  {Oxford University Press},\ \bibinfo {address} {New York, NY},\ \bibinfo
  {year} {1961})\BibitemShut {NoStop}%
\bibitem [{\citenamefont {{Trotta}}(2008)}]{2008ConPh..49...71T}%
  \BibitemOpen
  \bibfield  {author} {\bibinfo {author} {\bibfnamefont {R.}~\bibnamefont
  {{Trotta}}},\ }\href {\doibase 10.1080/00107510802066753} {\bibfield
  {journal} {\bibinfo  {journal} {Contemporary Physics}\ }\textbf {\bibinfo
  {volume} {49}},\ \bibinfo {pages} {71} (\bibinfo {year} {2008})},\ \Eprint
  {http://arxiv.org/abs/0803.4089} {arXiv:0803.4089} \BibitemShut {NoStop}%
\bibitem [{\citenamefont {{Kolb}}\ and\ \citenamefont
  {{Turner}}(1990)}]{kolb/turner:TEU}%
  \BibitemOpen
  \bibfield  {author} {\bibinfo {author} {\bibfnamefont {E.~W.}\ \bibnamefont
  {{Kolb}}}\ and\ \bibinfo {author} {\bibfnamefont {M.~S.}\ \bibnamefont
  {{Turner}}},\ }\href@noop {} {\emph {\bibinfo {title} {The Early Universe}}}\
  (\bibinfo  {publisher} {Addison-Wesley},\ \bibinfo {address} {New York, NY},\
  \bibinfo {year} {1990})\BibitemShut {NoStop}%
\bibitem [{\citenamefont {{Hinshaw}}\ \emph
  {et~al.}(2013{\natexlab{a}})\citenamefont {{Hinshaw}} \emph
  {et~al.}}]{hinshaw/etal:2013}%
  \BibitemOpen
  \bibfield  {author} {\bibinfo {author} {\bibfnamefont {G.}~\bibnamefont
  {{Hinshaw}}} \emph {et~al.},\ }\href {\doibase 10.1088/0067-0049/208/2/19}
  {\bibfield  {journal} {\bibinfo  {journal} {\apjs}\ }\textbf {\bibinfo
  {volume} {208}},\ \bibinfo {eid} {19} (\bibinfo {year}
  {2013}{\natexlab{a}})},\ \Eprint {http://arxiv.org/abs/1212.5226}
  {arXiv:1212.5226 [astro-ph.CO]} \BibitemShut {NoStop}%
\bibitem [{\citenamefont {{Planck
  Collaboration}}(2014{\natexlab{a}})}]{2013arXiv1303.5076P}%
  \BibitemOpen
  \bibfield  {author} {\bibinfo {author} {\bibnamefont {{Planck
  Collaboration}}},\ }\href {\doibase 10.1051/0004-6361/201321591} {\bibfield
  {journal} {\bibinfo  {journal} {\aap}\ }\textbf {\bibinfo {volume} {571}},\
  \bibinfo {eid} {A16} (\bibinfo {year} {2014}{\natexlab{a}})},\ \Eprint
  {http://arxiv.org/abs/1303.5076} {arXiv:1303.5076 [astro-ph.CO]} \BibitemShut
  {NoStop}%
\bibitem [{\citenamefont {Lyth}\ and\ \citenamefont
  {Riotto}(1999)}]{lyth/riotto:1999}%
  \BibitemOpen
  \bibfield  {author} {\bibinfo {author} {\bibfnamefont {D.~H.}\ \bibnamefont
  {Lyth}}\ and\ \bibinfo {author} {\bibfnamefont {A.}~\bibnamefont {Riotto}},\
  }\href@noop {} {\bibfield  {journal} {\bibinfo  {journal} {Phys. Rept.}\
  }\textbf {\bibinfo {volume} {314}},\ \bibinfo {pages} {1} (\bibinfo {year}
  {1999})}\BibitemShut {NoStop}%
\bibitem [{\citenamefont {{Chung}}\ \emph {et~al.}(2003)\citenamefont
  {{Chung}}, \citenamefont {{Shiu}},\ and\ \citenamefont
  {{Trodden}}}]{chung/shiu/trodden:2003}%
  \BibitemOpen
  \bibfield  {author} {\bibinfo {author} {\bibfnamefont {D.~J.}\ \bibnamefont
  {{Chung}}}, \bibinfo {author} {\bibfnamefont {G.}~\bibnamefont {{Shiu}}}, \
  and\ \bibinfo {author} {\bibfnamefont {M.}~\bibnamefont {{Trodden}}},\ }\href
  {\doibase 10.1103/PhysRevD.68.063501} {\bibfield  {journal} {\bibinfo
  {journal} {\prd}\ }\textbf {\bibinfo {volume} {68}},\ \bibinfo {eid} {063501}
  (\bibinfo {year} {2003})},\ \Eprint {http://arxiv.org/abs/astro-ph/0305193}
  {astro-ph/0305193} \BibitemShut {NoStop}%
\bibitem [{\citenamefont {{Ashoorioon}}\ \emph {et~al.}(2014)\citenamefont
  {{Ashoorioon}}, \citenamefont {{Dimopoulos}}, \citenamefont
  {{Sheikh-Jabbari}},\ and\ \citenamefont {{Shiu}}}]{2014arXiv1403.6099A}%
  \BibitemOpen
  \bibfield  {author} {\bibinfo {author} {\bibfnamefont {A.}~\bibnamefont
  {{Ashoorioon}}}, \bibinfo {author} {\bibfnamefont {K.}~\bibnamefont
  {{Dimopoulos}}}, \bibinfo {author} {\bibfnamefont {M.~M.}\ \bibnamefont
  {{Sheikh-Jabbari}}}, \ and\ \bibinfo {author} {\bibfnamefont
  {G.}~\bibnamefont {{Shiu}}},\ }\href {\doibase
  10.1016/j.physletb.2014.08.038} {\bibfield  {journal} {\bibinfo  {journal}
  {Physics Letters B}\ }\textbf {\bibinfo {volume} {737}},\ \bibinfo {pages}
  {98} (\bibinfo {year} {2014})},\ \Eprint {http://arxiv.org/abs/1403.6099}
  {arXiv:1403.6099 [hep-th]} \BibitemShut {NoStop}%
\bibitem [{\citenamefont {{Planck
  Collaboration}}(2014{\natexlab{b}})}]{2013arXiv1303.5082P}%
  \BibitemOpen
  \bibfield  {author} {\bibinfo {author} {\bibnamefont {{Planck
  Collaboration}}},\ }\href {\doibase 10.1051/0004-6361/201321569} {\bibfield
  {journal} {\bibinfo  {journal} {\aap}\ }\textbf {\bibinfo {volume} {571}},\
  \bibinfo {eid} {A22} (\bibinfo {year} {2014}{\natexlab{b}})},\ \Eprint
  {http://arxiv.org/abs/1303.5082} {arXiv:1303.5082 [astro-ph.CO]} \BibitemShut
  {NoStop}%
\bibitem [{\citenamefont {{Smith}}\ \emph {et~al.}(2014)\citenamefont
  {{Smith}}, \citenamefont {{Dvorkin}}, \citenamefont {{Boyle}}, \citenamefont
  {{Turok}}, \citenamefont {{Halpern}}, \citenamefont {{Hinshaw}},\ and\
  \citenamefont {{Gold}}}]{2014arXiv1404.0373S}%
  \BibitemOpen
  \bibfield  {author} {\bibinfo {author} {\bibfnamefont {K.~M.}\ \bibnamefont
  {{Smith}}}, \bibinfo {author} {\bibfnamefont {C.}~\bibnamefont {{Dvorkin}}},
  \bibinfo {author} {\bibfnamefont {L.}~\bibnamefont {{Boyle}}}, \bibinfo
  {author} {\bibfnamefont {N.}~\bibnamefont {{Turok}}}, \bibinfo {author}
  {\bibfnamefont {M.}~\bibnamefont {{Halpern}}}, \bibinfo {author}
  {\bibfnamefont {G.}~\bibnamefont {{Hinshaw}}}, \ and\ \bibinfo {author}
  {\bibfnamefont {B.}~\bibnamefont {{Gold}}},\ }\href {\doibase
  10.1103/PhysRevLett.113.031301} {\bibfield  {journal} {\bibinfo  {journal}
  {Physical Review Letters}\ }\textbf {\bibinfo {volume} {113}},\ \bibinfo
  {eid} {031301} (\bibinfo {year} {2014})},\ \Eprint
  {http://arxiv.org/abs/1404.0373} {arXiv:1404.0373} \BibitemShut {NoStop}%
\bibitem [{\citenamefont {{BICEP2 Collaboration}}(2014)}]{2014arXiv1403.3985B}%
  \BibitemOpen
  \bibfield  {author} {\bibinfo {author} {\bibnamefont {{BICEP2
  Collaboration}}},\ }\href {\doibase 10.1103/PhysRevLett.112.241101}
  {\bibfield  {journal} {\bibinfo  {journal} {Physical Review Letters}\
  }\textbf {\bibinfo {volume} {112}},\ \bibinfo {eid} {241101} (\bibinfo {year}
  {2014})},\ \Eprint {http://arxiv.org/abs/1403.3985} {arXiv:1403.3985}
  \BibitemShut {NoStop}%
\bibitem [{\citenamefont {{Starobinsky}}(1985)}]{starobinsky:1985}%
  \BibitemOpen
  \bibfield  {author} {\bibinfo {author} {\bibfnamefont {A.~A.}\ \bibnamefont
  {{Starobinsky}}},\ }\href@noop {} {\bibfield  {journal} {\bibinfo  {journal}
  {Soviet Astronomy Letters}\ }\textbf {\bibinfo {volume} {11}},\ \bibinfo
  {pages} {133} (\bibinfo {year} {1985})}\BibitemShut {NoStop}%
\bibitem [{\citenamefont {{Abazajian}}\ \emph {et~al.}(2014)\citenamefont
  {{Abazajian}}, \citenamefont {{Aslanyan}}, \citenamefont {{Easther}},\ and\
  \citenamefont {{Price}}}]{2014arXiv1403.5922A}%
  \BibitemOpen
  \bibfield  {author} {\bibinfo {author} {\bibfnamefont {K.~N.}\ \bibnamefont
  {{Abazajian}}}, \bibinfo {author} {\bibfnamefont {G.}~\bibnamefont
  {{Aslanyan}}}, \bibinfo {author} {\bibfnamefont {R.}~\bibnamefont
  {{Easther}}}, \ and\ \bibinfo {author} {\bibfnamefont {L.~C.}\ \bibnamefont
  {{Price}}},\ }\href {\doibase 10.1088/1475-7516/2014/08/053} {\bibfield
  {journal} {\bibinfo  {journal} {\jcap}\ }\textbf {\bibinfo {volume} {8}},\
  \bibinfo {eid} {053} (\bibinfo {year} {2014})},\ \Eprint
  {http://arxiv.org/abs/1403.5922} {arXiv:1403.5922} \BibitemShut {NoStop}%
\bibitem [{\citenamefont {{Hazra}}\ \emph
  {et~al.}(2014{\natexlab{a}})\citenamefont {{Hazra}}, \citenamefont
  {{Shafieloo}}, \citenamefont {{Smoot}},\ and\ \citenamefont
  {{Starobinsky}}}]{2014JCAP...06..061H}%
  \BibitemOpen
  \bibfield  {author} {\bibinfo {author} {\bibfnamefont {D.~K.}\ \bibnamefont
  {{Hazra}}}, \bibinfo {author} {\bibfnamefont {A.}~\bibnamefont
  {{Shafieloo}}}, \bibinfo {author} {\bibfnamefont {G.~F.}\ \bibnamefont
  {{Smoot}}}, \ and\ \bibinfo {author} {\bibfnamefont {A.~A.}\ \bibnamefont
  {{Starobinsky}}},\ }\href {\doibase 10.1088/1475-7516/2014/06/061} {\bibfield
   {journal} {\bibinfo  {journal} {\jcap}\ }\textbf {\bibinfo {volume} {6}},\
  \bibinfo {eid} {061} (\bibinfo {year} {2014}{\natexlab{a}})},\ \Eprint
  {http://arxiv.org/abs/1403.7786} {arXiv:1403.7786} \BibitemShut {NoStop}%
\bibitem [{\citenamefont {{Hazra}}\ \emph
  {et~al.}(2014{\natexlab{b}})\citenamefont {{Hazra}}, \citenamefont
  {{Shafieloo}}, \citenamefont {{Smoot}},\ and\ \citenamefont
  {{Starobinsky}}}]{2014arXiv1404.0360H}%
  \BibitemOpen
  \bibfield  {author} {\bibinfo {author} {\bibfnamefont {D.~K.}\ \bibnamefont
  {{Hazra}}}, \bibinfo {author} {\bibfnamefont {A.}~\bibnamefont
  {{Shafieloo}}}, \bibinfo {author} {\bibfnamefont {G.~F.}\ \bibnamefont
  {{Smoot}}}, \ and\ \bibinfo {author} {\bibfnamefont {A.~A.}\ \bibnamefont
  {{Starobinsky}}},\ }\href {\doibase 10.1103/PhysRevLett.113.071301}
  {\bibfield  {journal} {\bibinfo  {journal} {Physical Review Letters}\
  }\textbf {\bibinfo {volume} {113}},\ \bibinfo {eid} {071301} (\bibinfo {year}
  {2014}{\natexlab{b}})},\ \Eprint {http://arxiv.org/abs/1404.0360}
  {arXiv:1404.0360} \BibitemShut {NoStop}%
\bibitem [{\citenamefont {{Hazra}}\ \emph
  {et~al.}(2014{\natexlab{c}})\citenamefont {{Hazra}}, \citenamefont
  {{Shafieloo}}, \citenamefont {{Smoot}},\ and\ \citenamefont
  {{Starobinsky}}}]{2014arXiv1405.2012H}%
  \BibitemOpen
  \bibfield  {author} {\bibinfo {author} {\bibfnamefont {D.~K.}\ \bibnamefont
  {{Hazra}}}, \bibinfo {author} {\bibfnamefont {A.}~\bibnamefont
  {{Shafieloo}}}, \bibinfo {author} {\bibfnamefont {G.~F.}\ \bibnamefont
  {{Smoot}}}, \ and\ \bibinfo {author} {\bibfnamefont {A.~A.}\ \bibnamefont
  {{Starobinsky}}},\ }\href {\doibase 10.1088/1475-7516/2014/08/048} {\bibfield
   {journal} {\bibinfo  {journal} {\jcap}\ }\textbf {\bibinfo {volume} {8}},\
  \bibinfo {eid} {048} (\bibinfo {year} {2014}{\natexlab{c}})},\ \Eprint
  {http://arxiv.org/abs/1405.2012} {arXiv:1405.2012} \BibitemShut {NoStop}%
\bibitem [{\citenamefont {{Meerburg}}(2014)}]{2014arXiv1406.3243M}%
  \BibitemOpen
  \bibfield  {author} {\bibinfo {author} {\bibfnamefont {P.~D.}\ \bibnamefont
  {{Meerburg}}},\ }\href {\doibase 10.1103/PhysRevD.90.063529} {\bibfield
  {journal} {\bibinfo  {journal} {\prd}\ }\textbf {\bibinfo {volume} {90}},\
  \bibinfo {eid} {063529} (\bibinfo {year} {2014})},\ \Eprint
  {http://arxiv.org/abs/1406.3243} {arXiv:1406.3243} \BibitemShut {NoStop}%
\bibitem [{\citenamefont {{Kawasaki}}\ \emph {et~al.}(2014)\citenamefont
  {{Kawasaki}}, \citenamefont {{Sekiguchi}}, \citenamefont {{Takahashi}},\ and\
  \citenamefont {{Yokoyama}}}]{2014arXiv1404.2175K}%
  \BibitemOpen
  \bibfield  {author} {\bibinfo {author} {\bibfnamefont {M.}~\bibnamefont
  {{Kawasaki}}}, \bibinfo {author} {\bibfnamefont {T.}~\bibnamefont
  {{Sekiguchi}}}, \bibinfo {author} {\bibfnamefont {T.}~\bibnamefont
  {{Takahashi}}}, \ and\ \bibinfo {author} {\bibfnamefont {S.}~\bibnamefont
  {{Yokoyama}}},\ }\href {\doibase 10.1088/1475-7516/2014/08/043} {\bibfield
  {journal} {\bibinfo  {journal} {\jcap}\ }\textbf {\bibinfo {volume} {8}},\
  \bibinfo {eid} {043} (\bibinfo {year} {2014})},\ \Eprint
  {http://arxiv.org/abs/1404.2175} {arXiv:1404.2175} \BibitemShut {NoStop}%
\bibitem [{\citenamefont {{Planck
  Collaboration}}(2014{\natexlab{c}})}]{2013arXiv1303.5075P}%
  \BibitemOpen
  \bibfield  {author} {\bibinfo {author} {\bibnamefont {{Planck
  Collaboration}}},\ }\href {\doibase 10.1051/0004-6361/201321573} {\bibfield
  {journal} {\bibinfo  {journal} {\aap}\ }\textbf {\bibinfo {volume} {571}},\
  \bibinfo {eid} {A15} (\bibinfo {year} {2014}{\natexlab{c}})},\ \Eprint
  {http://arxiv.org/abs/1303.5075} {arXiv:1303.5075} \BibitemShut {NoStop}%
\bibitem [{\citenamefont {{Hinshaw}}\ \emph
  {et~al.}(2013{\natexlab{b}})\citenamefont {{Hinshaw}} \emph
  {et~al.}}]{2013ApJS..208...19H}%
  \BibitemOpen
  \bibfield  {author} {\bibinfo {author} {\bibfnamefont {G.}~\bibnamefont
  {{Hinshaw}}} \emph {et~al.},\ }\href {\doibase 10.1088/0067-0049/208/2/19}
  {\bibfield  {journal} {\bibinfo  {journal} {\apjs}\ }\textbf {\bibinfo
  {volume} {208}},\ \bibinfo {eid} {19} (\bibinfo {year}
  {2013}{\natexlab{b}})},\ \Eprint {http://arxiv.org/abs/1212.5226}
  {arXiv:1212.5226 [astro-ph.CO]} \BibitemShut {NoStop}%
\bibitem [{\citenamefont {{Anderson}}\ \emph {et~al.}(2012)\citenamefont
  {{Anderson}} \emph {et~al.}}]{2012MNRAS.427.3435A}%
  \BibitemOpen
  \bibfield  {author} {\bibinfo {author} {\bibfnamefont {L.}~\bibnamefont
  {{Anderson}}} \emph {et~al.},\ }\href {\doibase
  10.1111/j.1365-2966.2012.22066.x} {\bibfield  {journal} {\bibinfo  {journal}
  {\mnras}\ }\textbf {\bibinfo {volume} {427}},\ \bibinfo {pages} {3435}
  (\bibinfo {year} {2012})},\ \Eprint {http://arxiv.org/abs/1203.6594}
  {arXiv:1203.6594 [astro-ph.CO]} \BibitemShut {NoStop}%
\bibitem [{\citenamefont {{Padmanabhan}}\ \emph {et~al.}(2012)\citenamefont
  {{Padmanabhan}}, \citenamefont {{Xu}}, \citenamefont {{Eisenstein}},
  \citenamefont {{Scalzo}}, \citenamefont {{Cuesta}}, \citenamefont {{Mehta}},\
  and\ \citenamefont {{Kazin}}}]{2012MNRAS.427.2132P}%
  \BibitemOpen
  \bibfield  {author} {\bibinfo {author} {\bibfnamefont {N.}~\bibnamefont
  {{Padmanabhan}}}, \bibinfo {author} {\bibfnamefont {X.}~\bibnamefont {{Xu}}},
  \bibinfo {author} {\bibfnamefont {D.~J.}\ \bibnamefont {{Eisenstein}}},
  \bibinfo {author} {\bibfnamefont {R.}~\bibnamefont {{Scalzo}}}, \bibinfo
  {author} {\bibfnamefont {A.~J.}\ \bibnamefont {{Cuesta}}}, \bibinfo {author}
  {\bibfnamefont {K.~T.}\ \bibnamefont {{Mehta}}}, \ and\ \bibinfo {author}
  {\bibfnamefont {E.}~\bibnamefont {{Kazin}}},\ }\href {\doibase
  10.1111/j.1365-2966.2012.21888.x} {\bibfield  {journal} {\bibinfo  {journal}
  {\mnras}\ }\textbf {\bibinfo {volume} {427}},\ \bibinfo {pages} {2132}
  (\bibinfo {year} {2012})},\ \Eprint {http://arxiv.org/abs/1202.0090}
  {arXiv:1202.0090 [astro-ph.CO]} \BibitemShut {NoStop}%
\bibitem [{\citenamefont {{Beutler}}\ \emph {et~al.}(2011)\citenamefont
  {{Beutler}} \emph {et~al.}}]{2011MNRAS.416.3017B}%
  \BibitemOpen
  \bibfield  {author} {\bibinfo {author} {\bibfnamefont {F.}~\bibnamefont
  {{Beutler}}} \emph {et~al.},\ }\href {\doibase
  10.1111/j.1365-2966.2011.19250.x} {\bibfield  {journal} {\bibinfo  {journal}
  {\mnras}\ }\textbf {\bibinfo {volume} {416}},\ \bibinfo {pages} {3017}
  (\bibinfo {year} {2011})},\ \Eprint {http://arxiv.org/abs/1106.3366}
  {arXiv:1106.3366 [astro-ph.CO]} \BibitemShut {NoStop}%
\bibitem [{\citenamefont {{Feroz}}\ \emph {et~al.}(2009)\citenamefont
  {{Feroz}}, \citenamefont {{Hobson}},\ and\ \citenamefont
  {{Bridges}}}]{2009MNRAS.398.1601F}%
  \BibitemOpen
  \bibfield  {author} {\bibinfo {author} {\bibfnamefont {F.}~\bibnamefont
  {{Feroz}}}, \bibinfo {author} {\bibfnamefont {M.~P.}\ \bibnamefont
  {{Hobson}}}, \ and\ \bibinfo {author} {\bibfnamefont {M.}~\bibnamefont
  {{Bridges}}},\ }\href {\doibase 10.1111/j.1365-2966.2009.14548.x} {\bibfield
  {journal} {\bibinfo  {journal} {\mnras}\ }\textbf {\bibinfo {volume} {398}},\
  \bibinfo {pages} {1601} (\bibinfo {year} {2009})},\ \Eprint
  {http://arxiv.org/abs/0809.3437} {arXiv:0809.3437} \BibitemShut {NoStop}%
\bibitem [{\citenamefont {{Feroz}}\ \emph {et~al.}(2013)\citenamefont
  {{Feroz}}, \citenamefont {{Hobson}}, \citenamefont {{Cameron}},\ and\
  \citenamefont {{Pettitt}}}]{2013arXiv1306.2144F}%
  \BibitemOpen
  \bibfield  {author} {\bibinfo {author} {\bibfnamefont {F.}~\bibnamefont
  {{Feroz}}}, \bibinfo {author} {\bibfnamefont {M.~P.}\ \bibnamefont
  {{Hobson}}}, \bibinfo {author} {\bibfnamefont {E.}~\bibnamefont {{Cameron}}},
  \ and\ \bibinfo {author} {\bibfnamefont {A.~N.}\ \bibnamefont {{Pettitt}}},\
  }\href@noop {} {\bibfield  {journal} {\bibinfo  {journal} {ArXiv e-prints}\ }
  (\bibinfo {year} {2013})},\ \Eprint {http://arxiv.org/abs/1306.2144}
  {arXiv:1306.2144 [astro-ph.IM]} \BibitemShut {NoStop}%
\bibitem [{\citenamefont {{Lewis}}\ \emph {et~al.}(2000)\citenamefont
  {{Lewis}}, \citenamefont {{Challinor}},\ and\ \citenamefont
  {{Lasenby}}}]{Lewis:2000a}%
  \BibitemOpen
  \bibfield  {author} {\bibinfo {author} {\bibfnamefont {A.}~\bibnamefont
  {{Lewis}}}, \bibinfo {author} {\bibfnamefont {A.}~\bibnamefont
  {{Challinor}}}, \ and\ \bibinfo {author} {\bibfnamefont {A.}~\bibnamefont
  {{Lasenby}}},\ }\href {\doibase 10.1086/309179} {\bibfield  {journal}
  {\bibinfo  {journal} {\apj}\ }\textbf {\bibinfo {volume} {538}},\ \bibinfo
  {pages} {473} (\bibinfo {year} {2000})},\ \Eprint
  {http://arxiv.org/abs/arXiv:astro-ph/9911177} {arXiv:astro-ph/9911177}
  \BibitemShut {NoStop}%
\bibitem [{\citenamefont {{Lewis}}\ and\ \citenamefont
  {{Bridle}}(2002)}]{Lewis:2002a}%
  \BibitemOpen
  \bibfield  {author} {\bibinfo {author} {\bibfnamefont {A.}~\bibnamefont
  {{Lewis}}}\ and\ \bibinfo {author} {\bibfnamefont {S.}~\bibnamefont
  {{Bridle}}},\ }\href {\doibase 10.1103/PhysRevD.66.103511} {\bibfield
  {journal} {\bibinfo  {journal} {\prd}\ }\textbf {\bibinfo {volume} {66}},\
  \bibinfo {eid} {103511} (\bibinfo {year} {2002})},\ \Eprint
  {http://arxiv.org/abs/arXiv:astro-ph/0205436} {arXiv:astro-ph/0205436}
  \BibitemShut {NoStop}%
\bibitem [{\citenamefont {{Bastero-Gil}}\ \emph {et~al.}(2014)\citenamefont
  {{Bastero-Gil}}, \citenamefont {{Berera}}, \citenamefont {{Ramos}},\ and\
  \citenamefont {{Rosa}}}]{2014arXiv1404.4976B}%
  \BibitemOpen
  \bibfield  {author} {\bibinfo {author} {\bibfnamefont {M.}~\bibnamefont
  {{Bastero-Gil}}}, \bibinfo {author} {\bibfnamefont {A.}~\bibnamefont
  {{Berera}}}, \bibinfo {author} {\bibfnamefont {R.~O.}\ \bibnamefont
  {{Ramos}}}, \ and\ \bibinfo {author} {\bibfnamefont {J.~G.}\ \bibnamefont
  {{Rosa}}},\ }\href {\doibase 10.1088/1475-7516/2014/10/053} {\bibfield
  {journal} {\bibinfo  {journal} {\jcap}\ }\textbf {\bibinfo {volume} {10}},\
  \bibinfo {eid} {053} (\bibinfo {year} {2014})},\ \Eprint
  {http://arxiv.org/abs/1404.4976} {arXiv:1404.4976} \BibitemShut {NoStop}%
\bibitem [{\citenamefont {{Bousso}}\ \emph {et~al.}(2014)\citenamefont
  {{Bousso}}, \citenamefont {{Harlow}},\ and\ \citenamefont
  {{Senatore}}}]{2014arXiv1404.2278B}%
  \BibitemOpen
  \bibfield  {author} {\bibinfo {author} {\bibfnamefont {R.}~\bibnamefont
  {{Bousso}}}, \bibinfo {author} {\bibfnamefont {D.}~\bibnamefont {{Harlow}}},
  \ and\ \bibinfo {author} {\bibfnamefont {L.}~\bibnamefont {{Senatore}}},\
  }\href@noop {} {\bibfield  {journal} {\bibinfo  {journal} {ArXiv e-prints}\ }
  (\bibinfo {year} {2014})},\ \Eprint {http://arxiv.org/abs/1404.2278}
  {arXiv:1404.2278} \BibitemShut {NoStop}%
\bibitem [{\citenamefont {{Freivogel}}\ \emph {et~al.}(2014)\citenamefont
  {{Freivogel}}, \citenamefont {{Kleban}}, \citenamefont {{Rodriguez
  Martinez}},\ and\ \citenamefont {{Susskind}}}]{2014arXiv1404.2274F}%
  \BibitemOpen
  \bibfield  {author} {\bibinfo {author} {\bibfnamefont {B.}~\bibnamefont
  {{Freivogel}}}, \bibinfo {author} {\bibfnamefont {M.}~\bibnamefont
  {{Kleban}}}, \bibinfo {author} {\bibfnamefont {M.}~\bibnamefont {{Rodriguez
  Martinez}}}, \ and\ \bibinfo {author} {\bibfnamefont {L.}~\bibnamefont
  {{Susskind}}},\ }\href@noop {} {\bibfield  {journal} {\bibinfo  {journal}
  {ArXiv e-prints}\ } (\bibinfo {year} {2014})},\ \Eprint
  {http://arxiv.org/abs/1404.2274} {arXiv:1404.2274} \BibitemShut {NoStop}%
\bibitem [{\citenamefont {{Mortonson}}\ and\ \citenamefont
  {{Seljak}}(2014)}]{2014arXiv1405.5857M}%
  \BibitemOpen
  \bibfield  {author} {\bibinfo {author} {\bibfnamefont {M.~J.}\ \bibnamefont
  {{Mortonson}}}\ and\ \bibinfo {author} {\bibfnamefont {U.}~\bibnamefont
  {{Seljak}}},\ }\href {\doibase 10.1088/1475-7516/2014/10/035} {\bibfield
  {journal} {\bibinfo  {journal} {\jcap}\ }\textbf {\bibinfo {volume} {10}},\
  \bibinfo {eid} {035} (\bibinfo {year} {2014})},\ \Eprint
  {http://arxiv.org/abs/1405.5857} {arXiv:1405.5857} \BibitemShut {NoStop}%
\bibitem [{\citenamefont {{Flauger}}\ \emph {et~al.}(2014)\citenamefont
  {{Flauger}}, \citenamefont {{Hill}},\ and\ \citenamefont
  {{Spergel}}}]{2014arXiv1405.7351F}%
  \BibitemOpen
  \bibfield  {author} {\bibinfo {author} {\bibfnamefont {R.}~\bibnamefont
  {{Flauger}}}, \bibinfo {author} {\bibfnamefont {J.~C.}\ \bibnamefont
  {{Hill}}}, \ and\ \bibinfo {author} {\bibfnamefont {D.~N.}\ \bibnamefont
  {{Spergel}}},\ }\href {\doibase 10.1088/1475-7516/2014/08/039} {\bibfield
  {journal} {\bibinfo  {journal} {\jcap}\ }\textbf {\bibinfo {volume} {8}},\
  \bibinfo {eid} {039} (\bibinfo {year} {2014})},\ \Eprint
  {http://arxiv.org/abs/1405.7351} {arXiv:1405.7351} \BibitemShut {NoStop}%
\end{thebibliography}%

\label{lastpage}

\end{document}